**Common mistakes on "the production of electromagnetic waves" in many popular textbooks**

Fulin Zuo

Department of Physics

University of Miami

"**In the sense of our theory we more correctly represent the phenomenon by saying that fundamentally the waves which are being developed do not owe their formation solely to process at the origin, but arise out of the conditions of the whole surrounding space, which latter, according to our theory, is the true seat of the energy**." by **Heinrich Hertz (1893)**.

Electromagnetic wave is one of the most glorified subjects in physics with fascinating history of Michael Faraday,  James Maxwell and Heinrich Hertz and many others contributing to the field[1]. It is also one of the most difficult subjects in the college curricula even for many science and engineering majors due to the amount of mathematics involved and its abstract nature[2].  This makes it a difficult task to teach the materials effectively to the students, especially at the introductory college physics level.

It is even more difficult to convey the concept of the electromagnetic wave production and its propagation. It is no coincidence that it took more than 20 years after Maxwell proposed his theory of electromagnetic wave for Hertz to stumble accidentally on the discovery of the electromagnetic wave[1].

This may be related to the misleading or wrong presentations on the subject in many popular college physics textbooks adopted in thousands of college campuses worldwide. The consequence and gravity of common mistakes among Physics textbooks makes it a pressing issue to be investigated and corrected.

In several books we have available and surveyed [3-8], it has been found that the authors of the books use a very similar model to describe the wave propagation.  Electric field lines near the antenna are drawn from the positive charge to the negative charge as a function of the charge movement.  The completion of the charge movement corresponds to the cycle of the electric field wave propagating outward.  The corresponding magnetic field are usually not described in detail, but always given such that the direction of the cross product between the electric field and magnetic field points away from the dipole.

For example, a typical description of the process is like this:  *(a) At t=0 the electric field at point P is downward. (b) A short time later, the electric field at P is still downward, but now with a reduced magnitude. Note the field created at t=0 has moved to point Q. (c) After one-quarter of a cycle, at t=1/4T, the electric field at P vanishes. (d) The charge on the antenna has reversed polarity now, and the electric field at P points upward. (e) When the oscillator has completed half a cycle, t=1/2T, the field at point P is upward and of maximum magnitude. (f) At t=3/4T, the field at P vanishes again. The field produced at earlier times continues to move away from the antenna* (Quoted from James Walker[3]).

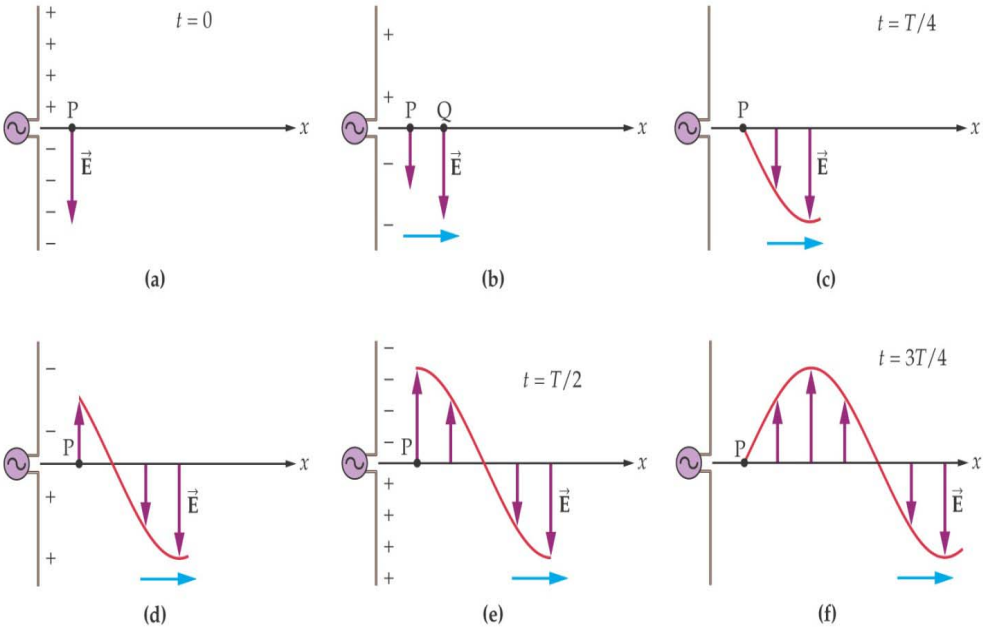

**Figure 1 From Walker Figure 25-1 producing an electromagnetic wave**

Similar descriptions can be found in many calculus based college physics books as well[4-8]. For example, in Tipler's *Physics for Scientist and Engineers* [4], very similar electric field diagram is used in the book (Figure 30-07, not shown here) and described like this: *At the time t=0, (Figure 30-07a), the ends of the rods are charged, and an electric field parallel to the rod exists near the rod. A magnetic field also exists, which is not shown, encircling the rods due to current in the rods.The fluctuations in these fields move out away from from the rods with the speed of light. After one fourth period, at t=1/4T(Figure 30-7b), the rods are uncharged, and the electric field near the rods is zero. At t=1/2T(Figure 30-7c), the rods are again charged, but the charges are opposite to those at t=0.*

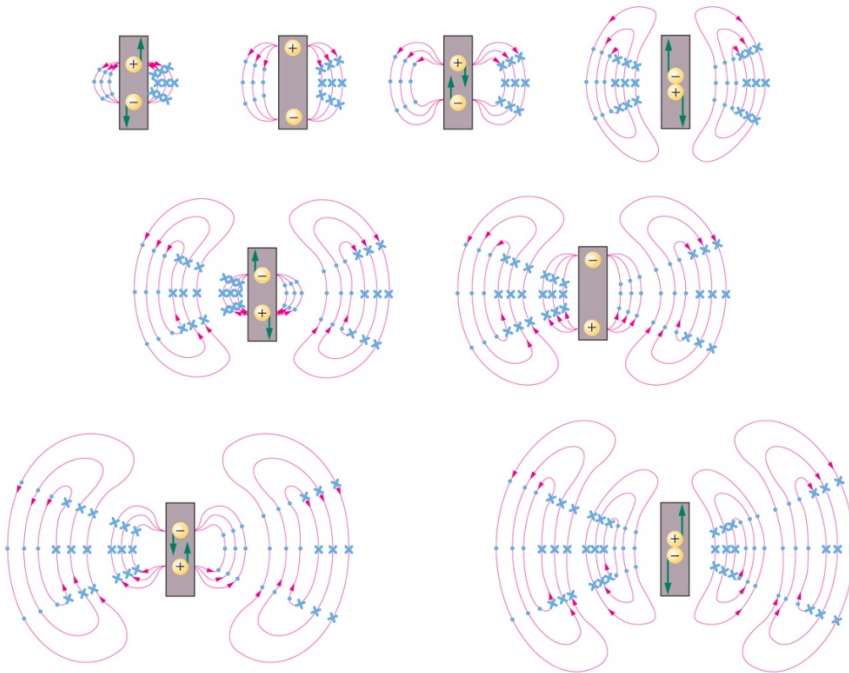

**Figure 2 From Tipler Figure 30-8**

In addition, the book plotted a

more detailed sequences of field lines with magnetic field included, as shown in Figure 2, with the following caption for the Figure: *"Electric field lines (in red) and magnetic field lines (in blue) produced by oscillating electric dipole. Each magnetic field line is a circle with the long axis of the dipole as its axis of revolution. The cross product EXB is directed from the dipole at all points"*[4].

The illustration and its description seem convenient and convincing enough to show that those waves are travelling outward and are the radiation waves. Many authors have adopted very similar description with small variations. The attractiveness of this approach lies in its simplicity and "apparent" plausibility.

The problem with this intuitive approach is that the fields illustrated in many of the books are not the radiation field, but the quasi-static field. Radiation field lines do not originate from or end on the charges!

The problem of dipole radiation is solved in many elementary Electricity and Magnetism books[2]. For a simple dipole along the z-axis $P(t) = q_o d \cos(\omega t) \hat{z}$, corresponding to a current $I(t) = -q_o \omega \sin \omega t$ the radiation electric field is given in spherical coordinates by[2]

$$E_r = -\frac{\mu_o P_o \omega^2}{4\pi} \left(\frac{\sin \theta}{r}\right) \cos[\omega(t - r/c)]\hat{\theta}$$

Radiation field dominates in the far field zone $(r \gg \frac{c}{\omega})$ and quasi-static field dominates in the near field zone $(r \leq \frac{c}{\omega})$ because of their dependence on $1/r$ and $1/r^3$, respectively. At a given point, the two fields coexist (in fact, an intermediate field $\propto 1/r^2$ exists as well but not important in the two zones). For example, in the near field zone, the quasi-static field $E_{qs} = \frac{P_o}{4\pi\epsilon_o}\left(\frac{\cos \omega(t-r/c)}{r^3}\right)(2 \cos \theta \, \hat{r} + \sin \theta \, \hat{\theta})$ is in $-\hat{z}$ direction on the y-axis while the radiation field $E_r$ points in the positive z direction near t=0 ($\hat{\theta}$ is $-\hat{z}$ on the y-axis). In fact, the radiation electric field is always pointing in the opposite direction to that of the quasi-static electric field along the y-axis. At point off-axis, the quasi-static field has finite radial component, while the radiating field has only component in $\hat{\theta}$ direction. And, of course, both fields vary periodically with its wavelength at a given time.

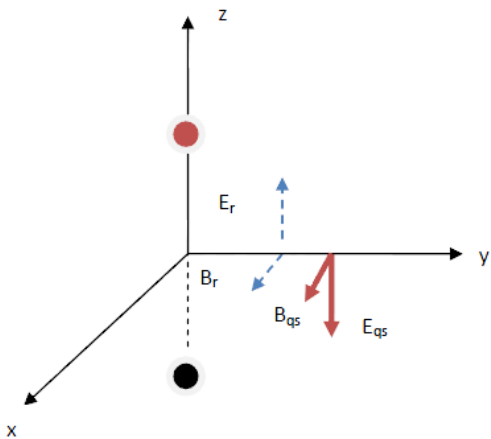

Figure 3 Electric and magnetic fields in the near zone at t=0

The radiation magnetic field in the oscillating dipole case is given by

$$B_r = -\frac{\mu_o P_o \omega^2}{4\pi c}\left(\frac{\sin \theta}{r}\right)\cos[\omega(t-r/c)]\hat{\phi}, \text{ and}$$

the corresponding quasi-static magnetic field:

$$B_{qs} = -\frac{\mu_o P_o \omega}{4\pi}\left(\frac{\sin \theta}{r^2}\right)\sin \omega(t-r/c)\,\hat{\phi}.$$

It is worth pointing out that in the near field zone where the quasi-static field dominates, the directions of the electric and magnetic field do not always generate the outgoing Poynting vector, as described for Figure 2 in the book[4]. Shown in the Figure 3 are the fields corresponding to the radiation and quasi-static electric and magnetic fields at t=0 in the near field zone. Note that the quasi-static field $E_{qs}$ points in the $-\hat{z}$ and the quasi-static magnetic field $B_{qs}$ in $\hat{x}$, resulting in a Poynting vector toward the dipole. The quasi-static field $E_{qs}$ and the radiating field $E_r$ are pointing in the opposite directions (They are not drawn to scale and are separated for clarity).

The radiating electric and magnetic fields are always in-phase with each other, resulting in a constant Poynting vector along the positive y direction on the y-axis. On the other hand, the direction of the cross product of the quasi-static electric and magnetic fields is given by $-\sin 2\omega(t - r/c)\,\hat{y}$, oscillating with both time and space. This is consistent with the fact that this is the quasi-static field, and no energy is transferred on average.

Based on the results of a simple dipole radiation, the intuitive presentation of the generation of the electromagnetic wave has the following problems: A) The electric field or field lines presented in many books[3-8] correspond to the quasi-static field in the near zone, and are not the radiation fields; B) The magnetic field is stated to always give rise to an outgoing wave, which is not true in the near zone; C) It fails to distinguish the radiation field from quasi-static field in different regions.

The presentations on the subject of radiation in many textbooks has been around for many decades, and appears plausible, but in fact misleading if not totally wrong. While it is a difficult subject to teach in great detail at the introductory level, it is important to emphasize the difference between the static field and the radiation field.

An illustrative picture of the radiation field component can be constructed by simply reversing the direction of the quasi-static electric field, as shown in Figure 4.

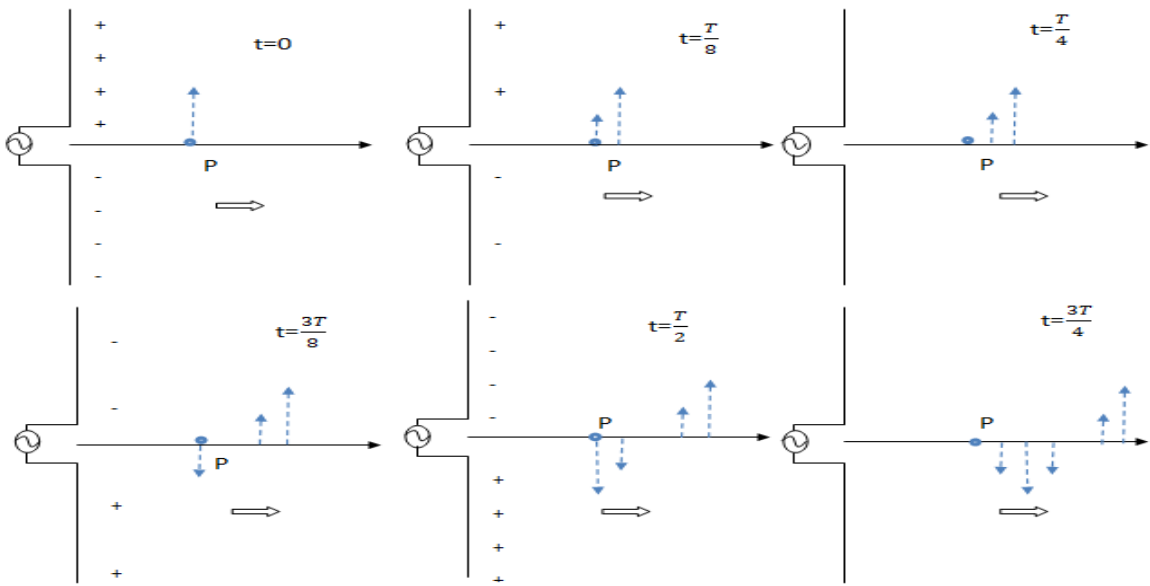

Figure 4 Radiation electric field as a function of time

The radiating magnetic field (not shown) is in phase with the electric field to give Poynting vector to the right.

It is important to stress the difference between the near field zone where static field dominates and the far field zone where radiation field propagates. The direction of the radiating electric field is counter intuitive because it is opposite to the quasi-static field.  Indeed, this can be used as a good example to emphasize the fact that electromagnetic wave is independent of charges, as described by Hertz in 1893[1].

The physics of dipole radiation is not new. However, it is surprising that the problem discussed here has not been addressed or corrected in the broad physics community as it is still permeated in many current college physics books[3-8].  It is true that most physics and some engineering majors will learn the rigorous treatment of radiation later on, but many others will not. Even though the consequence may be minimal for most students, it is paramount that as educators and physicists we should not teach our students wrong physics!

I want to thank Profs.  Arnold Perlmutter, Orlando Alvarez and especially Prof. Thom Curtright for very useful discussions.